\documentclass[fleqn,twoside]{article}
\usepackage{espcrc2,amssymb}

\def\oneloop {1L }
\def\twoloop {2L }

\def\bacol{\setlength{\arraycolsep}{0pt}}
\def\bec{\begin{center}}
\def\enc{\end{center}}
\def\[{$$}
\def\]{$$}
\def\ben{\begin{equation}}
\def\ba{\begin{array}}
\def\bea{\begin{eqnarray}}
\def\een{\end{equation}}
\def\eea{\end{eqnarray}}
\def\ea{\end{array}}
\def\btab{\begin{table}}
\def\btabu{\begin{tabular}}
\def\etab{\end{table}}
\def\etabu{\end{tabular}}
\def\bit{\begin{itemize}}
\def\eit{\end{itemize}}
\def\bef{\begin{figure}[htb]}
\def\befh{\begin{figure}[!h!]}
\def\enf{\end{figure}}

\def\tr{\mbox{Tr}}

\def\a{\alpha}

\def\hY{\hat{Y}}

\def\barU{{\bar{U}}}

\def\bmu{\mbox{\boldmath $\mu$}}

\def\g{\gamma}

\def\O{\Omega}

\def\S{\Sigma}

\def\Ga{\Gamma}
\def\GaR{{\Gamma_{\!\! R}}}

\def\half{{\textstyle{1 \over 2}}}

\def\b1{{\bf 1}}

\def\tA{\tilde{A}}

\def\bk{{\bf k}}
\def\bx{{\bf x}}
\def\by{{\bf y}}
\def\bz{{\bf z}}
\def\bp{{\bf p}}

\def\be{{\bf e}}
\def\b0{{\bf 0}}

\def\bv{{\bf v}}

\def\sbmu{{\mbox{\boldmath $\scriptstyle \mu$}}}

\def\barU{{\bar{U}}}

\def\sin{\hbox{sin}}

\def\nn{\nonumber}

\def\1{{1}}
\def\mod{\mbox{mod}}

\newcommand{\name}{\arabic{section}}
\newcommand{\newsection}[1]{\section{#1}\renewcommand{\theequation}
                              {\name.\arabic{equation}}
                            \setcounter{equation}{0}}
\newcommand{\newsubsection}[1]{\subsection{#1}\renewcommand{\theequation}
                               {\name.\arabic{subsection}.\arabic{equation}}
                               \setcounter{equation}{0}}

\def\rw{\rule[-5mm]{0mm}{12mm}}
\def\rw0{\rule[0mm]{0mm}{15mm}}
\def\rb{\raisebox{3mm}[0pt]}
\def\rb0{\raisebox{0mm}[0mm][20truemm]{\null}}

\usepackage{epsfig,latexsym}

\newlength{\figsize}
\newlength{\figoffset}
\newlength{\figbackup}
\newlength{\figendsp}

\def\preprints{
\vspace{-10ex}
{\small
\begin{tabbing}
\` Cambridge DAMTP-2002-103 \\
\` August 2002 \\
\end{tabbing} 
}
\vspace*{0.1in}
}

\title{
\preprints
Lattice perturbation theory for gluonic and fermionic actions.}

\author{I.T. Drummond, A. Hart, R.R. Horgan, L.C. Storoni%
        \address{{\small DAMTP, Cambridge University, 
          Wilberforce Road, Cambridge CB3 0WA, UK.}}
        }

\begin{document}

\begin{abstract}
\noindent
We calculate the two loop Landau mean links and the one loop
renormalisation of the anisotropy for Wilson and improved SU(3) gauge
actions, using twisted boundary conditions as a gauge invariant
infrared regulator. We show these accurately describe simulated
results, and outline a method for generating Feynman rules for general
lattice field theories, in a form suitable for efficient numerical
calculation of perturbative loop diagrams.
\end{abstract}

\maketitle

The conflicting demands of controlling finite volume effects and
regulating the computational effort are balanced by using a
large spatial lattice spacing, $a_s$. Symanzik improvement can be
employed to control discretisation effects. The temporal lattice
spacing, $a_t$, should remain small to give good resolution of
correlation functions and control the finite $a_t$ effects;
improvement in the temporal direction would lead to `ghost' poles in
the gluon propagator.  Whilst general directional asymmetries are best
couched in terms of a metric structure, we restrict ourselves here to
the case where only the temporal lattice spacing is reduced.

The anisotropy, $\chi \equiv a_s/a_t$, is renormalised. Physical
scales are related by the measured value, $\chi_R \equiv
Z(g^2,\chi) \chi$, which we calculate in perturbation theory
(PT). With tadpole improvement (using two loop (\twoloop)
factors derived in Sec.~\ref{sec_meanlink}), the one loop (\oneloop)
determination of $Z$ in Sec.~\ref{sec_anisotropy} is sufficient.


\newsection{\label{sec_python}\bf The perturbative action}

We follow the notation of L\"{u}scher and Weisz
\cite{luwe}.
The links of the cubical lattice in $D$ dimensions are
labeled $(\bx,\mu)$, $\mu = 1,...,D$ and the basis of the unit
cell is $\{ \be_\mu \}$.

The gauge field denoted $U$ is $U_\mu(\bv) = \exp(g
A_\mu(\bv+\half\be_\mu))$, where $g$ is the bare coupling constant. A
general action may be written as a sum of Schwinger lines made up of
links. In the gluonic case, these are closed contours. For a fermionic
action, the path is terminated at points $\bx$, $\by$ by quark fields
$\bar{\psi}$, $\psi$ respectively. The perturbative action and
associated vertex functions come from expanding the action as a
polynomial in $A$, {\it e.g.} in the gluonic case:
\[
\sum_r \frac{g^r}{r!}
\sum_{B_1 .. B_r} \tA^{a_1}_{\mu_1}(\bk_1) .. \tA^{a_r}_{\mu_r}(\bk_r)
V_r(B_1,..,B_r)
\]
with $B_i \equiv (\bk_i,\mu_i,a_i)$. The Feynman rule for the
$r$-point gluon vertex is $-V_r$. The concept is to use two separate
programs; one is used once to produce data files encoding $V_r$ from
the action expansion. The other then uses these data files many times
in evaluating loop integrals. Rather than store a very large lookup
table for all possible colour, Lorentz and momentum structure we focus
on the action--specific features. Writing $V_r(B_1,...,B_r) =
C_r(a_1,...,a_r) \; \hY_r(\bk_1,\mu_1 ; ... ; \bk_r,\mu_r)$, the
action independent colour factor is $C_r(a_1,...,a_r) = \tr ( T_{a_1}
... T_{a_r} ) + (-1)^r \tr ( T_{a_r} ... T_{a_1} )$ for a gluonic
action. We include it only in the loop integration code.  In addition
we can write for a given action $\hY_r(
\bmu_r;\bk_1,...,\bk_r) =$
\ben
\sum_{\a=1}^{n_r({\sbmu_r})} \frac{f_\a(\bmu_r)}{2r!}
\exp \left( i \sum_{i=1}^r \bk_i \cdot \bv^\a_i(\bmu_r) \right) .
\label{eqn_yvertex}
\een
For given Lorentz indices, $\bmu_r \equiv (\mu_1,...,\mu_r)$,
we have $n_r$ terms (clearly action dependent). Each term arises from
picking a sequence of $r$ gauge fields from the action with specific
Lorentz indices, and the positions of each gauge field are the
$\bv$'s. For simple actions, $n_r(\bmu_r) = 0$ for many $\bmu_r$.
So, for given $\bmu_r$, we can reconstruct $V_r$ for general momenta
and colour structure knowing only $( \{ f_\a ; v^\a_i,...,v^\a_r \} \;
| \; \a = 1,...,n_r )$.

\begin{figure}[t]
\leavevmode
\begin{center}
\vspace*{1ex}
\hbox{%
\epsfxsize = \figsize
\epsffile{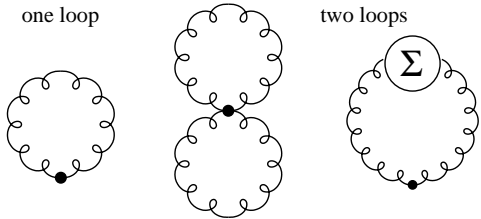}
}

\vspace{\figbackup}
\end{center}

\caption[]{\label{fig_meanlink_diags}\small Feynman diagrams for the
Landau mean link. $\Sigma$ refers to the diagrams in
Fig.~\ref{fig_anisotropy_diags}.}
\vspace{\figendsp}
\end{figure}

This must, however, be done $r!$ times, as we must consider all
permutations of the vertex legs in the loop integration code. We may
offset this cost by carrying out some of the symmetrisation at the
action expansion stage. This will be at the expense of increasing the
$n_r(\bmu_r)$ by the addition of permuted terms. We do not consider
colour factors in the action expansion, but a general permutation,
$\sigma$, will change $C_r$. We thus break the symmetrisation down
into two stages; the first comprises $\sigma$ that alter $C_r$ by a
factor independent of the colour index values. These are the cyclic
permutations and the inversion, where $\sigma \cdot C_r =
\chi_r(\sigma) C_r$, and can be included in the Taylor expansion code
by absorbing $\chi_r(\sigma)$ into the $f_\a$ of the permuted terms.
For $r \ge 4$, the remaining permutations are colour dependent and
must be carried out in the loop integration code.

We perform the Taylor expansion using a {\sc Python} code, chosen for
its object orientation and list handling properties. For a more
detailed specification of the implementation see
\cite{meanlink02}.
The Taylor expansion of the links in a path (or paths) is represented
by an instance of the {\tt taylor} class, $F$. This is implemented as
a `dictionary' whose keys are lists of Lorentz indices, $\bmu =
[\mu_1,...,\mu_r]$ for $r$ up to some chosen maximum order. $F[\bmu]$
returns a list of $n_r$ instances of the {\tt entity} class of order
$r$. Each encodes a monomial contribution to Eqn.~(\ref{eqn_yvertex})
and may be implemented as a list: $E_r = [ f,\bx,\by,\bv_1,...,\bv_r
]$, where $\bx$, $\by$ are redundant for gluonic actions. To
illustrate, for the single link $U_\mu(\bz)$:
\bea
F[[]] & = & [1,\bz,\bz + \be_\mu]
\nn \\
F[[\mu]] & = & [1,\bz,\bz + \be_\mu,\bz + \half \be_\mu]
\nn \\
F[[\mu,\mu]] & = & [1,\bz,\bz + \be_\mu,\bz + \half \be_\mu,
\bz + \half \be_\mu]
\nn
\eea
{\it etc.} It is useful to store position vector components as
(integer) multiples of $a/2$. We demand that the product of two {\tt
taylor}s representing two paths should be the Taylor expansion of the
combined contour obtained by following the first path, then the
second. Gauge covariance demands that the second {\tt taylor} be thus
translated to start where the first path ceased. In the above example,
$F^\prime = F * F$ is the expansion of $U_\mu(\bz) U_\mu(\bz +
\be_\mu)$, and $F^\prime * F^\prime$ would be a four link
object. This allows for compact definitions of actions using templates
that are themselves composite, or recursive, functions of other
objects. In particular, covariant derivatives and fat links can be
described without explicitly expanding all the contributing paths.

For a pure gauge action, as well as being translationally invariant,
the final paths are closed and traced, and with only the real part
taken. This high degree of symmetry permits identification of many
equivalent terms in $F$, and a significant compression of the {\tt
taylor} can be carried out prior to symmetrisation.

\begin{figure}[t]
\leavevmode
\begin{center}
\vspace*{1ex}
\hbox{%
\epsfxsize = \figsize
\epsffile{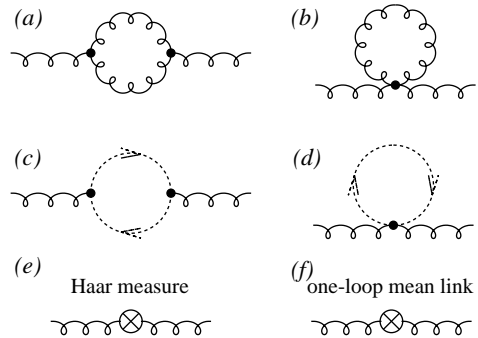}
}

\vspace{\figbackup}
\end{center}

\caption[]{\label{fig_anisotropy_diags}\small
Feynman diagrams for the self energy}
\vspace{\figendsp}
\end{figure}
\newsubsection{\label{prop} the propagator}

The gluon propagator is computed for given momentum by inverting the
two-point function $V^{(2)}_{\mu\nu}$ having fixed the gauge by adding
$V^{(2)}_{\mu\nu}$ an extra term.  The inverse propagator is then
\[
\Ga^{(\a)}_{\mu\nu}(\bk) = V^{(2)}_{\mu\nu}(\bk) + 
\hat{k}_\mu \hat{k}_\nu / (\a \chi) \; ,
\]
where $\hat{k}_\mu \equiv 2 \; \sin (k_\mu/2)$ and $\alpha=1$ is
Feynman gauge. Inverting this, a general gauge $\g$ (such as Landau
gauge, $\g=0$) has a propagator (up to a colour factor)
\[
G^{(\g)}_{\mu\nu}(\bk) = G^{(\a)}_{\mu\nu}(\bk) + 
\frac{ (\a-\g) \chi \hat{k}_\mu \hat{k}_\nu}
{(\chi^2 {\hat{k}_0}^2 + {\hat{k}_i}^2)} \; .
\]

\begin{figure}[t]
\leavevmode
\begin{center}
\vspace*{1ex}
\hbox{%
\hspace{\figoffset}
\epsfxsize = \figsize
\epsffile{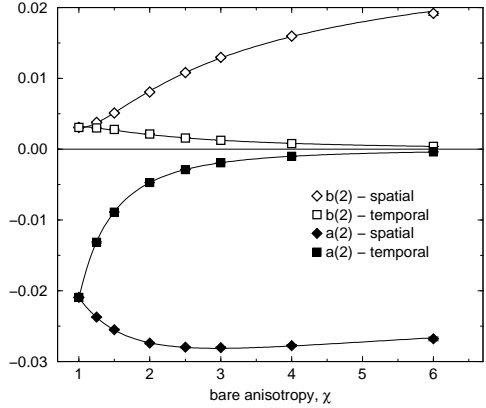}
}

\vspace{\figbackup}
\end{center}

\caption[]{\label{W2L}\small
\twoloop Landau mean link coefficients for the Wilson action.}
\vspace{\figendsp}
\end{figure}

We follow 
\cite{luwe} 
and use twisted periodic boundary conditions for the gauge
field. There is no zero mode and no concomitant infrared
divergences whilst gauge invariance is still maintained.

The twisted boundary condition for gauge fields is $U_\mu(\bx+L_\nu
\be_\nu) = \O_\nu U_\mu(\bx)\O^{-1}_\nu$, where the twist matrices
$\O_\nu$ are constant SU(N) matrices which satisfy $\O_\mu\O_\nu =
z_{\mu\nu}\O_\nu\O_\mu$ with $z_{\mu\nu}=\exp(2 \pi i n_{\mu\nu}/N)$
an element of the centre of SU(N) with $n_{\mu\nu} = 0,1,..., N-1$, an
antisymmetric tensor. In the calculations presented below we use
`doubly twisted' boundary conditions: $n_{12}=-n_{21} = 1$, $n_{\mu\nu}
= 0$ otherwise, and also `quadruply twisted': $n_{ij}\epsilon_{ijk} =
(1,1,1),\;n_{0j} = (1,1,1)$. The former is used as in
\cite{luwe} 
to define a physical gluon mode and the latter most effectively
suppresses tunnelling between equivalent vacua and so allows the
simulation of the Landau meanlink to match PT.  It is
important that $n_{ij}n_{0k}\epsilon_{ijk} = 0\;\mod{3}$ so that the
instanton number is zero
\cite{gonzalez97}.

\begin{figure}[t]
\begin{center}
\vspace*{1ex}
\hbox{%
\hspace{\figoffset}
\epsfxsize = \figsize
\epsffile{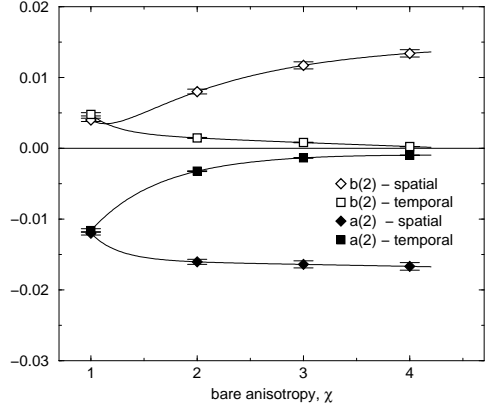}
}

\vspace{\figbackup}
\end{center}

\caption[]{\label{S2L}\small 
\twoloop Landau mean link coefficients 
for the SI action.}
\vspace{\figendsp}
\end{figure}

For the double twist with lattice of extent $L$ in the $1,2$
directions and continous in $3,4$ the momentum spectrum is continuous
in $k_3,k_4$ and discrete in $k_1,k_2$ with
\[
k_i = \frac{2 \pi \bar{k}_i}{L} + \frac{2\pi n_i}{NL}
\]
$i=1,2$, and $\bar{k}_i, n_i \in \mathbb{Z}$. The interstitial
momenta, $n_i$, play the role of the colour index and excluding
$n_1=n_2=0$ eliminates the zero mode and imposes a gauge-invariant
infrared cutoff momentum of $2\pi/NL$. The Clebsch-Gordan factor is
now a complex phase
\cite{luwe}.
When a graph is built it must have the same overall centre phase
factor as the tree diagram of the same external topology. For example,
the self energy must have the same factor as the inverse propagator.

\begin{table*}

\def\none	{\multicolumn{2}{c}{---}}

\caption{\label{tab_expansions}\small
Fits to perturbative expansion coefficients as functions of $\chi$, as
defined in the text.}

\begin{center}
\begin{tabular}{ccr@{.}lr@{.}lr@{.}lr@{.}l}
\hline
action & quantity & 
\multicolumn{2}{c}{constant} & 
\multicolumn{2}{c}{$1 / \chi$} & 
\multicolumn{2}{c}{$1 / \chi^2$} & 
\multicolumn{2}{c}{$\log (\chi) / \chi$} \\
\hline
Wilson & $b_s^{(1)}$ &
$-0$&01232 ~ (2) & 0&0931 ~ (1) & 
$-0$&048 ~ (1) & $-0$&049 ~ (1) \\
& $b_t^{(1)}$ &
\none & 0&09132 ~ (2) & 
0&01385 ~ (9) & 0&0090 ~ (1) \\
\cline{2-10}
 & $a_s^{(2)}$ &
$-0$&017 ~ (1) & 0&019 ~ (3) & 
$-0$&023 ~ (4) & $-0$&041 ~ (4) \\
 & $a_t^{(2)}$ &
\none & 0&0036 ~ (2) & 
$-0$&0246 ~ (2) & $-0$&011 ~ (1) \\
 & $b_s^{(2)}$ &
0&030 ~ (1) & $-0$&042 ~ (3) & 
0&015 ~ (4) & $-0$&013 ~ (4) \\
 & $b_t^{(2)}$ &
\none & 0&0103 ~ (2) & 
$-0$&0072 ~ (2) & $-0$&038 ~ (1) \\
\cline{2-10}
 & $\eta$ &
0&1687 ~ (2) & $-0$&16397 ~ (4) & 
$-0$&005245 ~ (2) & \none \\
\hline
Symanzik & $b_s^{(1)}$ &
$-0$&1012 ~ (2) & 0&0895 ~ (4) & 
$-0$&0513 ~ (6) & $-0$&0502 ~ (7) \\
improved & $b_t^{(1)}$ &
\none & $-0$&0194 ~ (1) & 
$-0$&0453 ~ (1) & 0&0110 ~ (1) \\
& $b_s^{(2)}$ &
0&0091017 & $-0$&071095 & 
0&065993 & 0&051785 \\
& $b_t^{(2)}$ &
$-0$&007214 & $-0$&016691 & 
0&028623 & 0&028351 \\
\cline{2-10}
 & $\eta$ &
0&0955 ~ (4) & $-0$&0702 ~ (16) & 
$-0$&0399 ~ (14) & \none \\
 & $\eta^\prime$ &
0&0602 ~ (1) & $-0$&0656 ~ (2) & 
$-0$&0237 ~ (1) & \none \\
\hline
\end{tabular}
\end{center}

\vspace{\figendsp}
\end{table*}

The Feynman rules arising from the ghost fields used in the
Fadeev--Popov gauge fixing, and from the measure are given in
\cite{drummond02}.

%
%
%
%
%

We consider the Wilson action (W) and the Symanzik improved action
(SI) in
\cite{alea},
{\bacol
\bea
S_W && (\beta,\chi) = 
- \beta \left( \chi P_{s,t} + \frac{1}{\chi} P_{s,s'} \right) 
\nn \\
S_{SI} && (\beta,\chi,v) = 
- \beta \left( \chi \left\{ \frac{4}{3} P_{s,t}
- \frac{1}{12} \frac{R_{ss,t}}{v^2} \right\} \right.
\nn \\
+ && \left. \frac{1}{\chi} \left\{ \frac{5}{3} P_{s,s'} 
 - \frac{1}{12} 
\left[ \frac{R_{ss,s'}}{v^2} + \frac{R_{s's',s}}{v^2} \right] 
\right \} \right)
\nn
\eea
}
summing over spatial directions $s$, $s^\prime < s$. $P$ are
plaquettes, and $R$ denotes $2 \times 1$ loops. We use these forms of
the actions as they minimise the dependence of the action parameters
on the coupling, $g^2 = 6/\beta$.

Tadpole improvement factors, $u_{s,t}$, arising from chosen
self-consistency conditions amount to a reparametrisation of $S_W$
\[
\beta = \frac{\beta_0}{{u_s}^3 u_t} \equiv \frac{6}{{g_0}^2} \; ,
~~ 
\chi = \frac{\chi_0 u_s}{u_t} \; .
\]
In the SI case, we must also impose $v = u_s$, and PT is specific to a
particular improvement scheme. If $u_s = 1 + d_s g^2 + O(g^4)$, then
{\bacol
\bea
S_{SI} && (\beta,\chi,v) = S_{SI}(\beta,\chi,v=1) + g^2 \Delta S_{SI}
\nn \\
\Delta S_{SI} && = \textstyle\frac{\beta d_s}{6} 
\left( \chi R_{ss,t} + \chi^{-1} (R_{ss,s'} + R_{s's',s}) \right)
\nn
\eea
}
and we treat $g^2 \Delta S_{SI}$ as a counterterm insertion
(Fig.~\ref{fig_anisotropy_diags}(f)) in the gluon propagator.

\newsection{\label{sec_meanlink} Landau mean link improvement}

We focus on the Landau mean link scheme, where $u_{s,t}$
are defined as the expectation values of spatial and temporal links in
Landau gauge.

With $b^{(1)}_l = a^{(1)}_l$, $b^{(2)}_l = a^{(2)}_l + a^{(1)}_l
(3a^{(1)}_s+a^{(1)}_t)$, the series for $u_{l=s,t}$ can be written as 
{\bacol
\[
u_l  = \left\{ 
\begin{array}{lll}
1 + a^{(1)}_l g^2     & + a^{(2)}_l g^4     & + O(g^6) \; , \\ 
1 + b^{(1)}_l {g_0}^2 & + b^{(2)}_l {g_0}^4 & + O({g_0}^6) \; .
\end{array}
\right.
\]
}
The relevant diagrams are shown in Fig.\ref{fig_meanlink_diags}.  The
\twoloop contributions were computed using a parallel version of {\sc
Vegas}.  Fits to the $L \to \infty$ \oneloop and \twoloop
contributions are given in Table~\ref{tab_expansions} and shown in
Figs.~\ref{W2L},~\ref{S2L}.

By comparing such truncated series with results from high-$\beta$
Monte Carlo simulations, higher order coefficients in the perturbative
expansion may be inferred
\cite{trottier01}.
Field configurations were generated using a 2nd order Runge-Kutta
Langevin updating routine. The implementation of the Langevin
evolution is such that any pure gauge action can be simulated by
simply specifying the list of paths and the associated couplings. The
group derivative for each loop is then computed automatically by an
iterative procedure which moves around the loop link by link
constructing the appropriate traceless anti-Hermitian contribution to
the Langevin velocity. This is the most efficient implementation,
minimising the number of group multiplications needed. It applies
whenever the quantity to be differentiated is specified as a Wilson
path.

The twisted boundary conditions are implemented in the manner
suggested in
\cite{luwe} 
where the field simulated, $\barU = U$ everywhere, save
$U_\mu(\bx)\O_\mu$ when $x_\mu = L_\mu$, where $\O_\mu$ is the twist
matrix associated with the $\mu$ direction and $L_\mu$ is the lattice
extent. The action $S(\barU)$ is identical to the untwisted action
except that a loop whose projection onto the $(\mu,\nu)$-plane,
$\mu<\nu$, encircles the point $(L_\mu+\half,L_\nu+\half)$ has an
additional factor of $(z_{\mu
\nu})^{-c}$ where $c$ is the integer winding number. The program
assigns the correct phase factors to each path appropriately. A
Fourier accelerated algorithm is vital when fixing to Landau gauge,
which is otherwise prohibitively time consuming.

\begin{figure}[t]
\leavevmode
\begin{center}
\vspace*{1ex}
\hbox{%
\hspace{\figoffset}
\epsfxsize = \figsize
\epsffile{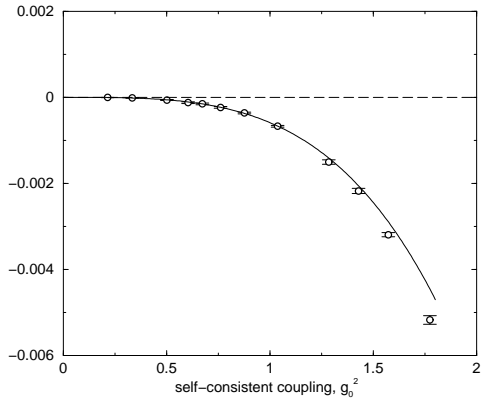}
}

\vspace{\figbackup}
\end{center}

\caption[]{\label{Wsim_l}\small 
Deviation of the Landau mean link from \twoloop
PT as a function of ${g_0}^2$ for the Wilson action.}
\vspace{\figendsp}
\end{figure}

Both the mean link and plaquette values are measured for a range of
$\beta$ and $\chi$ values.  In Fig.~\ref{Wsim_l} we plot for the
isotropic Wilson action the simulation results (with \twoloop PT
already subtracted) versus ${g_0}^2$, showing the deviation from the
\twoloop PT.  Care must be taken when fitting polynomials to data in a
given range of the expansion parameter. If the true coefficients are
such that in this finite window there is an approximate cancellation
between some combination of terms, then {\em all} such terms must be
included in the fit function.  If not, as might be the case when
trying polynomials of increasing order, marked instability in goodness
of fit and even low order coefficients will be seen. Whilst this did
not seem to be the case here, we fit the data points using `Baysian'
techniques
\cite{lepage01}
to obtain $b^{(3)}_s = -0.00032 \; (5)$, $b^{(4)}_s = -0.00027 \;
(5)$.  We believe that the value of $b^{(3)}_s$ is reliable and
$b^{(4)}_s$ is plausible.


\newsection{\label{sec_anisotropy} The renormalised anisotropy}

\begin{figure}[t]
\leavevmode
\begin{center}
\vspace*{1ex}
\hbox{%
\hspace{\figoffset}
\epsfxsize = \figsize
\epsffile{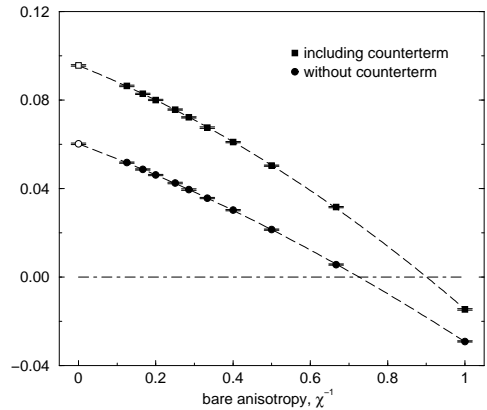}
}

\vspace{\figbackup}
\end{center}

\caption[]{\label{SIeta}\small The \oneloop anisotropy renormalization
coefficient versus anisotropy for the Symanzik action.}
\vspace{\figendsp}
\end{figure}

We calculate $Z(g^2,\chi) = 1 + \eta(\chi) g^2$ to \oneloop from the
dispersion relation for the gluon propagator in twisted QCD
\cite{luwe}.

The calculations were done on lattices with $8L^2I^2$ twisted momentum
grid points, where $4 \le L \le 32$ is the lattice size in the twisted
$1,2$-directions and $I > 50$ gives an effectively continuous momentum
spectrum in the untwisted $0,3$-directions. The gluon mode has
momentum parametrised by $\bp = (ip_0,m,0,p_3) \; ,~~ m = {2\pi/NL}$,
where $m$ is the pole mass of the bare gluon propagator. We calculate
the pole energy of the propagator as a function of $p_3$. For
sufficiently small $p_3$ the dispersion relation can be fitted to the
standard quadratic form using the renormalised mass, $m_R$, and
$\chi_R$ as parameters. The dispersion relation is derived from
\[
\GaR^{(\a)}_{\mu\nu}(\bp) = \Ga^{(\a)}_{\mu\nu}(\bp) - 
g^2\S_{\mu\nu}(\bp) = 0 \; ,
\]
To $O(g^2)$ the momentum argument of $\S_{\mu\nu}(\bp)$ is given by
the zero of the bare inverse propagator, most conveniently done for
$\mu=2$.
So, for given $L,$ $p_3$, we determine the bare pole value
$p_0=E_0$ by numerical solution of $\Ga^{(\a)}_{2\,2}(\bp) = 0$,
and the bare mass $m_0$ is defined by $E_0 = m_0/\chi$ at $p_3=0$.

The lattice energy function is determined by the inverse propagator,
$F(E) = 4\sinh^2(E/2)$. We define $\chi_R$ using dispersion
relations for very small $p_3$
\bea
F(E_0) & = & p_3^2 / \chi^2 +   F(m_0 / \chi) \; ,
\nn \\
F(E_R) & = & p_3^2 / \chi_R^2 + F(m_R / \chi_R) \; ,
\nn
\eea
with $E_R$ defined by
\[
\chi F (E_0) - \chi F(E_R) = g^2 \S(E_0,m,0,p_3) \; .
\]
To $O(g^2)$ we define $m_R = m_0 + g^2 m_1$ and
\[
\eta \left[ \frac{p_3^2}{\chi^2} + 
\frac{m_0}{2\chi} F^\prime \left( \frac{m_0}{\chi} \right) \right] -
\frac{m_1}{2\chi} F^\prime \left( \frac{m_0}{\chi} \right) = 
\frac{\Sigma}{2\chi} \; .
\]
$\eta(L)$ and $m_1(L)$ are determined by a straight line
fit, and extrapolated $L \to \infty$.

\begin{figure}[t]
\leavevmode
\begin{center}
\vspace*{1ex}
\hbox{%
\hspace{\figoffset}
\epsfxsize = \figsize
\epsffile{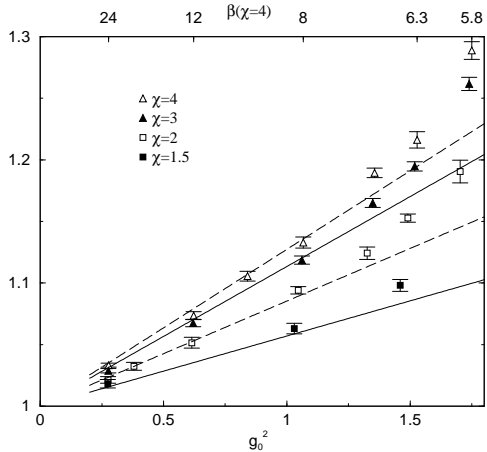}
}

\vspace{\figbackup}
\end{center}

\caption[]{\label{fig_klassen}\small
Comparing $Z$ from simulation
\cite{klassen98}
and \oneloop PT for the Wilson action.}
\vspace{\figendsp}
\end{figure}

The \oneloop diagrams for the self-energy $\S_{\mu\nu}(\bp)$ are shown
in Fig.~\ref{fig_anisotropy_diags}. For on-shell external momentum
$(iE_0,m,0,p_3)$, graphs Fig.~\ref{fig_anisotropy_diags}(a,c) exhibit
a pole in the integrand so the $k_0$ integration contour is shifted
$k_0 \rightarrow k_0 - iE_0/2$. We also use the change of
variables
\cite{luwe}
$\bk \to \bk^\prime$ and $k_\mu = k_\mu^\prime - \alpha_\mu
\sin(k_\mu^\prime)$. A good choice is $\a_\mu \sim 1-(\chi
L_\mu)^{-1}$. The integrals did not require {\sc Vegas} and took
between 2 and 16 hours on a PC. Feynman gauge was used.

For the Wilson action we verify $\eta(\chi=1)=1 ~ \forall ~ L$. For
$L \to \infty$, the mass renormalisation $m_1/m_0$ agreed with
\cite{snip}, 
and $\eta(\chi > 1)$ with
\cite{peba},
both derived using the background field method. In
Fig.~\ref{fig_klassen} we compare the \oneloop PT prediction for $Z$
with measurements from
\cite{klassen98}.
The \twoloop mean link expressions from Table~\ref{tab_expansions}
have been used to rescale the coupling. The error in $\chi_R$ is $<
5\%$ for $\beta \ge 5.8$ and $< 2 \%$ for $\beta \ge 6.3$ for $1.5 \le
\chi \le 6$. This is partly due to the error in the expressions for
$u_{s,t}$, {\it e.g.} Fig.~\ref{Wsim_l}, and the truncation error in
$\chi_R$ is expected to be smaller.

Using the same method, for the SI action we consider actions with mean
link improvement ($d_s = b_s^{(1)}$) and without ($d_s = 0$ and
omitting Fig.~\ref{fig_anisotropy_diags}(f)), obtaining $\eta$ and
$\eta^\prime$ respectively. These are shown in Fig.~\ref{SIeta},
and fits to $\chi$ are given in Table~\ref{tab_expansions}. In
Table~\ref{chis} we compare these predictions to measured data from
\cite{alea,shigemitsu02}.
The gauge couplings simulated are relatively small.  For $\chi_0 \le
4$ the error on $\chi_R$ is 2\%, and for $\chi_0 \le 6$ only 3\%. As
the simulated values of $u_{s,t}$ were used to rescale
$\chi_0$,~$\beta_0$, these errors reflect the \oneloop truncation
error on $Z({g_0}^2,\chi$).

\oneloop calculations of $Z$ thus seem to be sufficiently accurate for
use in setting the scale {\it etc.} in most lattice applications.

\begin{table}
\caption[]{\label{chis}\small
Comparing $\chi_R$ from simulation
\cite{alea,shigemitsu02}
and \oneloop PT for the SI action.}
\bec
\begin{tabular}{ccccc}
\hline
$\beta_0$ & $\chi_0$ & 
\multicolumn{1}{c}{$a_s^{-1}/\mathrm{MeV}$} & 
\multicolumn{1}{c}{$\chi_R^\mathrm{(meas.)}$} &
\multicolumn{1}{c}{$\chi_R^\mathrm{(pert.)}$} \\
\hline
1.7 &  4 & 661 (11) & 
3.56 (2) & 3.601 \\
    &  6 & 779 (28) &
5.28 (2) & 5.463 \\
\hline
1.8 &  4 & 797 (21) & 
3.61 (2) & 3.629 \\ 
    &  6 & 839 (9) & 
5.31 (2) & 5.489 \\
\hline
2.4 &  3 & 1200 (50) & 
2.71 (3) & 2.776 \\
\hline
\end{tabular}
\enc
\vspace{\figendsp}
\end{table}

We are pleased to acknowledge the use of the Hitachi SR2201 at the
University of Tokyo Computing Centre and the Cambridge--Cranfield High
Performance Computing Facility for this work.

{\small
\vspace{-2ex}

}


\begin{thebibliography}{10}

\bibitem{luwe}
M.~L\"{u}scher and P.~Weisz,
\newblock Nucl. Phys. {\bf B266}, 309 (1986).

\bibitem{meanlink02}
I.~Drummond {\em et~al.},
\newblock in preparation  (2002).

\bibitem{gonzalez97}
A.~Gonzalez-Arroyo,
\newblock hep-th/9807108.

\bibitem{drummond02}
I.~T. Drummond, A.~Hart, R.~R. Horgan and L.~C. Storoni,
\newblock hep-lat/0208010.

\bibitem{alea}
M.~Alford {\em et~al.},
\newblock Phys. Rev. {\bf D63}, 074501 (2001), [hep-lat/0003019].

\bibitem{trottier01}
H.~D. Trottier {\em et~al.},
\newblock Phys. Rev. {\bf D65}, 094502 (2002), [hep-lat/0111028].

\bibitem{klassen98}
T.~R. Klassen,
\newblock Nucl. Phys. {\bf B533}, 557 (1998), [hep-lat/9803010].

\bibitem{lepage01}
G.~P. Lepage {\em et~al.},
\newblock Nucl. Phys. Proc. Suppl. {\bf 106}, 12 (2002), [hep-lat/0110175].

\bibitem{shigemitsu02}
J.~Shigemitsu {\em et~al.},
\newblock hep-lat/0207011.

\bibitem{snip}
J.~Snippe,
\newblock Nucl. Phys. {\bf B498}, 347 (1997), [hep-lat/9701002].

\bibitem{peba}
M.~Garcia~Perez and P.~van Baal,
\newblock Phys. Lett. {\bf B392}, 163 (1997), [hep-lat/9610036].

\end{thebibliography}
\end{document}